\documentclass[aps,prl,superscriptaddress,amsmath,amssymb,twocolumn,a4paper,floatfix]{revtex4-1}
\usepackage{graphicx}
\usepackage{float}
\usepackage{hyperref}
\usepackage{bm}
\usepackage{tabularx}
\usepackage{xcolor}
\usepackage{bbm}
\usepackage{accents}

\newcolumntype{Z}{>{\centering\let\newline\\\arraybackslash\hspace{0pt}}X}

\begin{document}
\title{Spin and Charge Correlations across the Metal-to-Insulator Crossover in the Half-Filled 2D Hubbard model}
\author{Aaram J. Kim}
\email{aaram.kim@kcl.ac.uk}
\affiliation{Department of Physics, King's College London, Strand, London WC2R 2LS, UK}

\author{Fedor Simkovic IV}
\altaffiliation[current address: ]{Centre de Physique Th\'eorique, \'Ecole Polytechnique,
CNRS, Universit\'e Paris-Saclay, 91128 Palaiseau, France and Coll\`{e}ge de France, 11 place Marcelin Berthelot, 75005 Paris, France.}

\affiliation{Department of Physics, King's College London, Strand, London WC2R 2LS, UK}

\author{Evgeny Kozik}
\email{evgeny.kozik@kcl.ac.uk}
\affiliation{Department of Physics, King's College London, Strand, London WC2R 2LS, UK}

\begin{abstract}
The 2D Hubbard model with nearest-neighbor hopping on the square lattice and an average of one electron per site is known to undergo an extended crossover from metallic to insulating behavior driven by proliferating antiferromagnetic correlations. We study signatures of this crossover in spin and charge correlation functions and present results obtained with controlled accuracy using the diagrammatic Monte Carlo approach in the range of parameters amenable to experimental verification with ultracold atoms in optical lattices. The qualitative changes in charge and spin correlations associated with the crossover are observed at well-separated temperature scales, which encase the intermediary regime of non-Fermi-liquid character, where local magnetic moments are formed and nonlocal fluctuations in both channels are essential. 
\end{abstract}

\maketitle 

Recent developments of quantum emulators based on ultracold atoms loaded in an optical lattice~\cite{Bloch:2005uv,Lewenstein:2007hr,Greif:2015bg,Hart2015, Cheuk:2016kq,Mazurenko:2017ec,Brown:2017dy, Nichols:2019iq} have enabled accurate experimental realization and probing of the quintessential single-band 2D Hubbard model of correlated electrons in solids: 
\begin{equation}
	\mathcal{H} = -t \!\! \sum^{}_{\langle \mathbf{xy}\rangle,\sigma} \!\! (c^{\dagger}_{\mathbf{x}\sigma}c^{}_{\mathbf{y}\sigma} + \mathrm{H.c.}) 
+ U\sum^{}_{\mathbf{x}}n_{\mathbf{x} \uparrow} n_{\mathbf{x} \downarrow} - \mu\sum^{}_{\mathbf{x}\sigma}n_{\mathbf{x}\sigma},
	\label{eqn:Hamiltonian}
\end{equation}
where $c^{}_{\mathbf{x}\sigma}$ annihilates a fermion with spin $\sigma$ on the site $\mathbf{x}$, $\langle \mathbf{xy}\rangle$ implies nearest-neighbor sites, $n_{\sigma}(\mathbf{x}) = c^{\dagger}_{\mathbf{x}\sigma}c^{}_{\mathbf{x}\sigma}$ is the corresponding number operator, 
$t$ is the hopping amplitude (set to unity), $U$ the on-site repulsion, and $\mu$ the chemical potential.
Despite seeming simplicity, the model harbors extremely rich physics, including, e.g., unconventional~\cite{deng2015emergent} and possibly high-temperature superconductivity~\cite{anderson1997theory}, while \textit{a priori} accurate theoretical results in the thermodynamic limit are remarkably scarce~\cite{benchmarks}. 

Central among properties of the Hubbard model is the state of the interaction-induced insulator at half-filling ($\langle n_{\uparrow} + n_{\downarrow} \rangle=1$), when the noninteracting system is a metal. An important ingredient is the tendency toward antiferromagnetic (AFM) ordering due to nesting of the Fermi surface (FS), i.e. the existence of a single wave vector $\mathbf{Q}=(\pi,\pi)$ that connects any point on the FS to another point on the FS. At $U/t \ll 1$, an exponentially small $\sim t \exp(-2\pi \sqrt{t/U})$ energy gap in charge excitations emerges due to an exponential increase of the AFM correlation length~\cite{Rev:1951ib}, despite the absence of long-range order at any $T>0$~\cite{Mermin:1966da,Hohenberg:1967br}. At $U/t \gg 1$, the charge gap $\sim U/2$ is due to on-site repulsion, while AFM correlations develop at much smaller scales $\sim 4t^2/U$ and are irrelevant for the insulator. This drastic qualitative difference between the limiting cases---a local scenario at strong coupling versus that local in the momentum space at weak coupling---makes physics at intermediate $U \sim t$ particularly intriguing and challenging to describe reliably.      


When extended AFM correlations are explicitly suppressed, a Mott insulator is expected to emerge by a first-order metal-to-insulator transition~\cite{Georges:1996zz,Maier:2005tj,Moukouri:2001ch,Kyung:2003ju,Gull:2008fw,Park:2008ec,2009PhRvB..80x5102G,Werner:2009ck,Gull:2013hh,Fratino:2017jw,Walsh:2019uk}. In the 2D Hubbard model (\ref{eqn:Hamiltonian}) currently realized in experiments, extending correlations make the insulator develop in a smooth crossover~\cite{Lichtenstein2000afm_sc, Schafer:2015jg, Rohringer:2016jt, Fedor:2018vo}. Current quantum emulators~\cite{Greif:2015bg,Cheuk:2016kq,Mazurenko:2017ec,Brown:2017dy,Nichols:2019iq} have already reached the range of temperatures of the crossover. The structure of spin correlations can be measured with single-site resolution~\cite{Parsons:fp, Mazurenko:2017ec}. Compressibility and non-local density fluctuations can be directly probed~\cite{Duarte2015_compressiblity, drewes2016_density_fluctuations}. These techniques provide a powerful toolset to pinpoint the location of the crossover and characterize the underlying mechanisms, for which reliable theoretical results are currently missing. In a broader context, the role of AFM correlations in non-Fermi-liquid (nFL) physics has been the subject of extensive research and is widely believed to be relevant for unconventional superconductivity~\cite{Lee:2006de}. Despite the absence of the fermionic sign problem at half filling, accurate description of the crossover has proven to be extremely challenging for Monte Carlo methods limited to finite-size systems due to the substantial size dependence~\cite{Schafer:2015jg, Fedor:2018vo}. Recent controlled results~\cite{Fedor:2018vo} by the diagrammatic determinant Monte Carlo algorithm for the self-energy ($\Sigma$DDMC)~\cite{SimkovicIV:2017tn}, which works directly in the thermodynamic limit (TDL), and the large-cluster dynamical cluster approximation~\cite{leblanc:2013} demonstrate that the crossover is nontrivial and involves a transitional nFL \footnote{The metallic behavior of the half-filled 2D Hubbard model is, strictly speaking, not of the conventional Fermi-liquid type either~\cite{Afchain2005}, but we do not make this distinction here.} regime with a partially gapped FS~\cite{Schafer:2015jg, Rohringer:2016jt}.   

\begin{figure}[]
	\centering
	\includegraphics[width=\columnwidth]{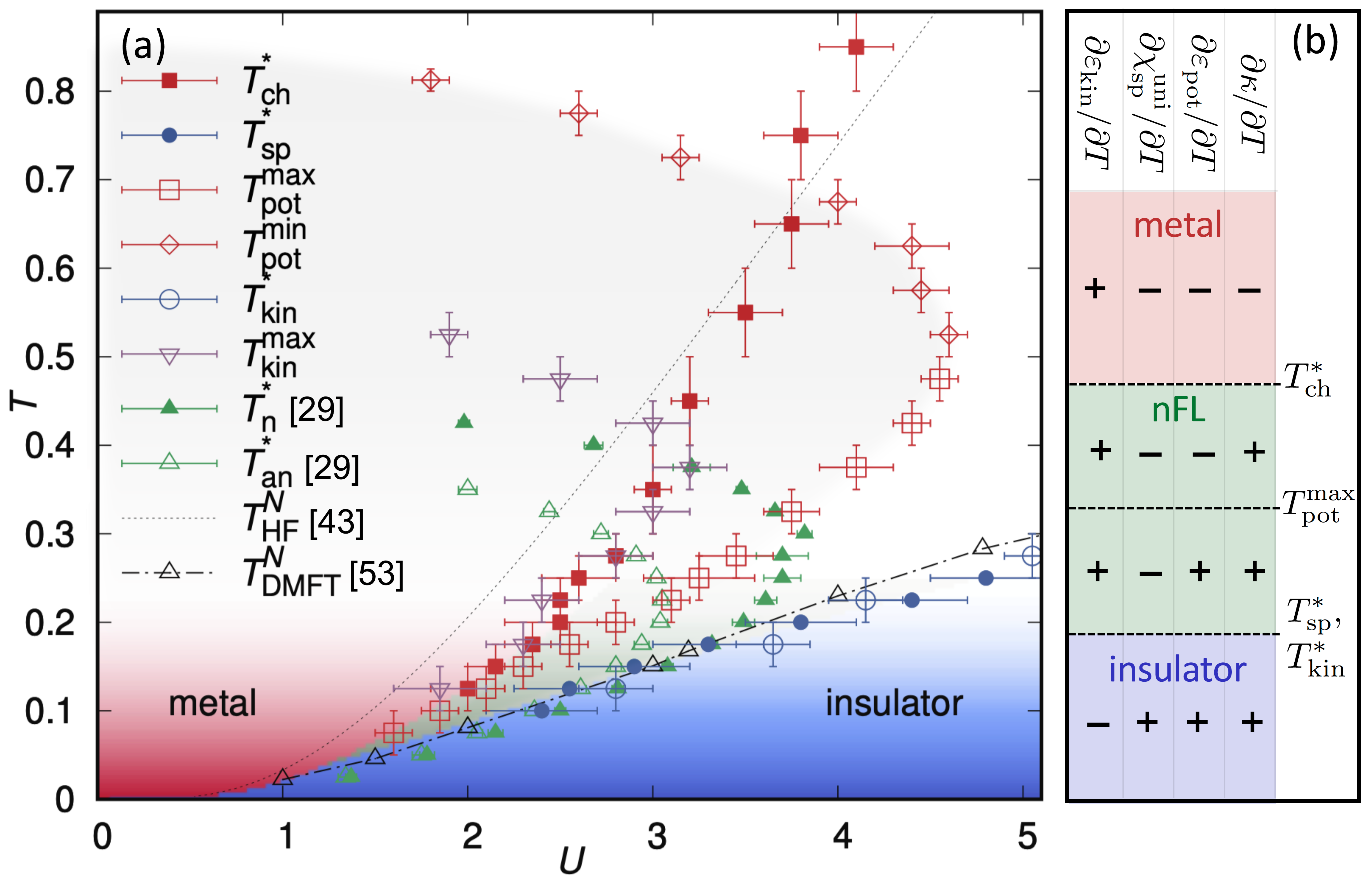}
	\caption{
		(a) Diagram of the extended metal-to-quasi-AFM-insulator crossover (see text). (b) Schematic of sign changes of the observables defining the corresponding crossover temperatures.  
	}
	\label{fig:phasediagram}
\end{figure}

In this Letter, we study with controlled accuracy experimentally observable signatures of the metal-insulator crossover in the equal-time spin and charge correlation functions as well as in potential and kinetic energies. We employ the connected determinant diagrammatic Monte Carlo (CDet) algorithm~\cite{Rossi:2017kpa} in the TDL and the approach of Ref.~\cite{SimkovicIV:2017tn} for controlled evaluation of observables from their diagrammatic series in the strongly correlated regime. The results are summarized in Fig.~\ref{fig:phasediagram}. Crossover temperatures are defined as the points where the derivatives with respect to $T$ of the compressibility $\kappa$ ($T^*_\mathrm{ch}$), uniform spin susceptibility $\chi_\mathrm{sp}^\mathrm{uni}$ ($T^*_\mathrm{sp}$), potential energy $\varepsilon_\mathrm{pot}$---or double occupancy $\langle d \rangle = \langle n_{\uparrow}n_{\downarrow}\rangle$, $\varepsilon_\mathrm{pot}=U\langle d \rangle$--- ($T^{\mathrm{max,min}}_\mathrm{pot}$), and kinetic energy $\varepsilon_\mathrm{kin}$ ($T^*_\mathrm{kin}$) change their signs. At relevant couplings, $\langle d \rangle(T)$ exhibits a maximum (at $T^{\mathrm{max}}_\mathrm{pot}$) and a minimum (at $T^{\mathrm{min}}_\mathrm{pot}$), and $\frac{d \varepsilon_\mathrm{kin}}{d T}$ has a maximum as a function of $U$ at $T^{\mathrm{max}}_\mathrm{kin}$. Below $T^*_\mathrm{an}$, $T^*_\mathrm{n}$, obtained by $\Sigma$DDMC in Ref.~\cite{Fedor:2018vo}, the self-energy becomes manifestly insulatorlike---its imaginary part is lowest at the lowest Matsubara frequency---at the antinodal $\mathbf{k}=(\pi,0)$ and nodal $\mathbf{k}=(\pi/2,\pi/2)$ momentum points respectively. At $T \lesssim 0.25$, where quasiparticle properties (and thus the notions of metal, nFL, and insulator) become meaningful~\cite{Fedor:2018vo}, nFL behavior is observed in the region between $T^*_\mathrm{ch}(U)$ and $T^*_\mathrm{sp}(U)$ (green shading in Fig.~\ref{fig:phasediagram}). Upon cooling, $\kappa$ becomes insulator-like first at $T^*_\mathrm{ch}$, while $\varepsilon_\mathrm{pot}$, $\varepsilon_\mathrm{kin}$, and $\chi_\mathrm{sp}^\mathrm{uni}$ are still of metallic character [as summarized in Fig.~\ref{fig:phasediagram}(b)], the AFM correlation length $\xi$ at strong coupling ($U\sim 3$) is only as long as about two lattice constants, and the self-energy does not yet exhibit insulating behavior anywhere on the FS. Long-range AFM correlations with $\xi \gtrsim 10$ develop at a notably lower $T^*_\mathrm{sp}$, below which all studied observables are insulatorlike. Nonlocal fluctuations are key for the existence of the transitional nFL, while the changes observed in this regime, such as restructuring of spatial correlations and development of the local magnetic moment, enable the crossover and generically require a range of parameters to take place. 

In the diagrammatic Monte Carlo approach to the Hubbard model~\cite{VanHoucke:2010ky,Kozik:2010fla} one computes numerically exactly the coefficients of the Taylor-series expansion in $U$ for a given observable in the TDL. For an order-$N_V$ coefficient of a two-body correlation function this amounts to summing all connected Feynman diagrams with four fixed external vertices $i$, $i'$, $o$, $o'$, the number $N_V$ of internal vertices $V$, and integrating over all configurations of $V$. The CDet~\cite{Rossi:2017kpa} algorithm allows us to evaluate the integrand at an exponential cost using determinantal summation of a factorial number of diagram topologies~\cite{rubtsov2003, Rubtsov2005determinants} with a recursive subtraction of all disconnected diagrams, while the integration over $V$ can be subsequently performed by Monte Carlo sampling~\cite{rubtsov2003, Rubtsov2005determinants, Bourovski:2004jf, Rossi:2017kpa,SimkovicIV:2017tn,Moutenet:2018ib}. For a given configuration $i,i', o, o', V$, the sum of all diagram topologies $a_{ii'oo'}(V)$ is obtained as a determinant of a 
matrix constructed from noninteracting Green's functions~\cite{rubtsov2003}. The sum of all connected diagrams $c_{ii'o o'}(V)$ can be found by a recursive subtraction of disconnected topologies following Ref.~\cite{Rossi:2017kpa},
\begin{eqnarray}
	c_{ii'oo'}(V) &=& a_{ii'oo'}(V) - \sum^{}_{S\subsetneq V}c_{ii'oo'}(S)a_{\emptyset}(V\setminus S) \nonumber\\
	&&- \sum^{}_{S\subset V}c_{io}(S)c_{i'o'}(V\setminus S),
	\label{eqn:recursion}
\end{eqnarray}
where $a_{\emptyset}(V)$ is the determinantal sum of all closed-loop diagrams without open ends and $c_{io}(V)$ is that of connected diagrams with only two external vertices. The last term becomes relevant when the variable whose correlations are computed has a finite expectation value. 
In the regime of interest, the series are convergent; obtaining their coefficients with the statistical error bar $ \lesssim 10\%$ at the highest accessible orders $N_V=9-11$ allows us to reliably evaluate the corresponding observables by a controlled extrapolation to infinite order~\cite{SimkovicIV:2017tn, Fedor:2018vo}.

\begin{figure}[t]
	\centering
	\includegraphics[width=0.45\textwidth]{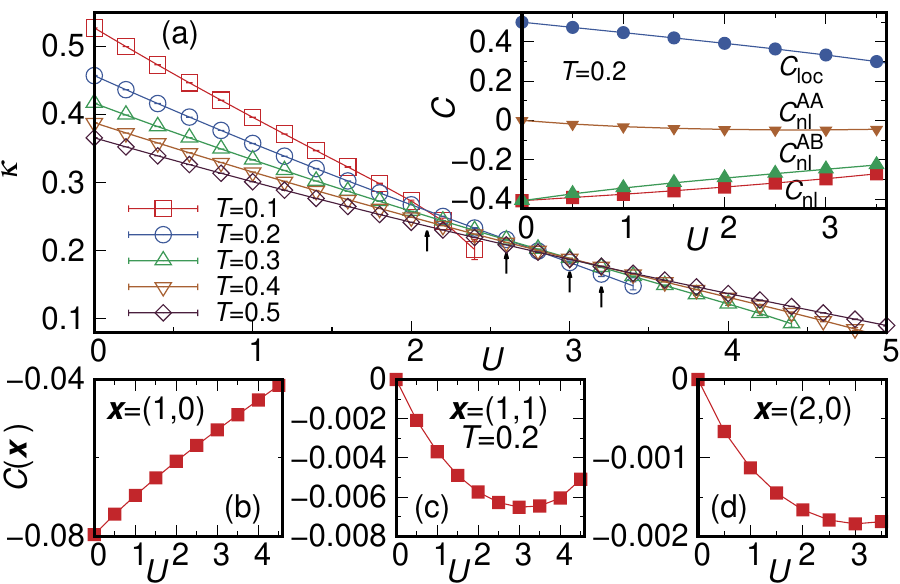}
	\caption{
		(a) Compressibility $\kappa$ versus $U$ for various temperatures. Arrows mark the crossings between $\kappa(U)$ at consecutive $T$. Inset: net contributions to $\kappa$ from density correlations: local ($C_\mathrm{loc}$), nonlocal within the same ($C_{\mathrm{nl}}^{\mathrm{AA}}$) and different ($C_{\mathrm{nl}}^{\mathrm{AB}}$) sublattices.  
		(b)-(d) Charge correlator Eq.~(\ref{eqn:Nx}) at $T=0.2$ for (b) $\mathbf{x}=(1,0)$, (c) $\mathbf{x}=(1,1)$, and (d) $\mathbf{x}=(2,0)$.
	}
	\label{fig:kappa_nl}
\end{figure}

The equal-time density-density (charge) correlator 
\begin{equation}
	C(\mathbf{x}-\mathbf{y}) = \langle \delta n(\mathbf{x})\delta n(\mathbf{y})\rangle,
	\label{eqn:Nx}
\end{equation}
where $\delta n(\mathbf{x}) = \sum_\sigma n_\sigma(\mathbf{x}) - \langle n_\sigma\rangle$, provides a direct signature of insulating behavior via the compressibility $\kappa = \frac{\partial n}{\partial \mu} =  \frac{\beta}{N}\sum^{}_{\mathbf{x},\mathbf{y}}C(\mathbf{x}-\mathbf{y})$,
with $N$ the number of lattice sites. Figure~\ref{fig:kappa_nl}(a) shows $\kappa$ as a function of $U$ at various temperatures. The temperature dependence of $\kappa$ gives an indication of the character of the system. In the metallic regime at small $T$ and $U$, $\kappa \propto -\ln T$ is dominated by the van Hove divergence of the density of states on the FS, so that $\partial\kappa/\partial T \propto - 1/T$ is negative. At large $U$, the system is an insulator with a charge gap and temperature-activated density fluctuations, so that $\partial\kappa/\partial T$ is positive. The condition $\partial\kappa/\partial T=0$, satisfied at the crossings of consecutive curves in Fig.~\ref{fig:kappa_nl}(a), thus defines the crossover scale $T_{\mathrm{ch}}^*(U)$. 
It is noticeably higher than $T^*_\mathrm{an}$, suggesting that the criterion of an nFL based on emergence of a polelike feature in the self-energy is a strong condition. At low $T$, $T_{\mathrm{ch}}^*(U)$ follows qualitatively, albeit systematically lower, the N\'{e}el temperature of the Hartree-Fock approximation $T^N_{\mathrm{HF}}$~\cite{Borejsza:2003gu}.
At $U,T \gg t$, it approaches its atomic limit asymptote $T_{\mathrm{ch}}^*(U) \approx 0.3911 U$. 

The local part of the charge correlator $C_\mathrm{loc}=C(0)$ is related to double occupancy, $C_\mathrm{loc}=2 \langle d \rangle$. Although the monotonic decrease of $\kappa$ with $U$ is mainly driven by the suppression of $\langle d \rangle$, $T_{\mathrm{ch}}^*$ is substantially enhanced by nonlocal fluctuations described by $C_{\mathrm{nl}}=\frac{1}{N}\sum_{\mathbf{x} \neq \mathbf{y}} C(\mathbf{x}-\mathbf{y})$ near the crossover. Expressing compressibility as $\kappa=\beta (C_{\mathrm{loc}} + C_{\mathrm{nl}})$ we find that $\partial [\beta C_{\mathrm{loc}}]/ \partial T=0$ at a lower temperature $\approx T^*_\mathrm{sp}(U)$, i.e. nonlocal charge fluctuations lead to the separation between $T^*_\mathrm{ch}$ and $T^*_\mathrm{sp}$. The inset of Fig.~\ref{fig:kappa_nl}(a), where $C_{\mathrm{loc}}$ and $C_{\mathrm{nl}}$ are plotted versus $U$ at $T=0.2$, shows that nonlocal correlations are of the same order of magnitude as the local ones but different in sign, suggesting that the behavior of $\kappa$ follows from a delicate interplay between the two~\cite{Walsh:2019uk}.

As $U$ is increased, the nature of nonlocal charge correlations changes, Figs.~\ref{fig:kappa_nl}(b)-\ref{fig:kappa_nl}(d). At weak coupling, a short-range anticorrelation, the so-called Pauli suppression, of density fluctuations between lattice sites on different sublattices [shown for $\mathbf{x}=(1,0)$ in Fig.~\ref{fig:kappa_nl}(b)] originates from the fermionic statistics. As $U$ is increased, an anti-correlation within the same sublattice [$\mathbf{x}=(1,1), (2,0)$ in Figs.~\ref{fig:kappa_nl}(c-d)] develops, while that between different sublattices is continuously suppressed. This so-called correlation hole, exclusively due to interactions, forms rapidly at small $U$~\cite{Gorelik:2012cx}. However, below $T^*_\mathrm{ch}(U)$ $|C(\mathbf{x}=(1,1), (2,0))|$ starts saturating and is eventually suppressed at larger $U$ in the insulating regime. The nFL is therefore marked by a broad minimum of nonlocal charge correlations between the sites on the same sublattice. The net contributions to $C_{\mathrm{nl}}$ from $\mathbf{x}$ and $\mathbf{y}$ belonging to the same ($C_{\mathrm{nl}}^{\mathrm{AA}}$) and different ($C_{\mathrm{nl}}^{\mathrm{AB}}$) sublattices are plotted in the inset of Fig.~\ref{fig:kappa_nl}(a), where it is seen that their dependence on $U$ is qualitatively different.

\begin{figure}[]
	\centering
	\includegraphics[width=0.45\textwidth]{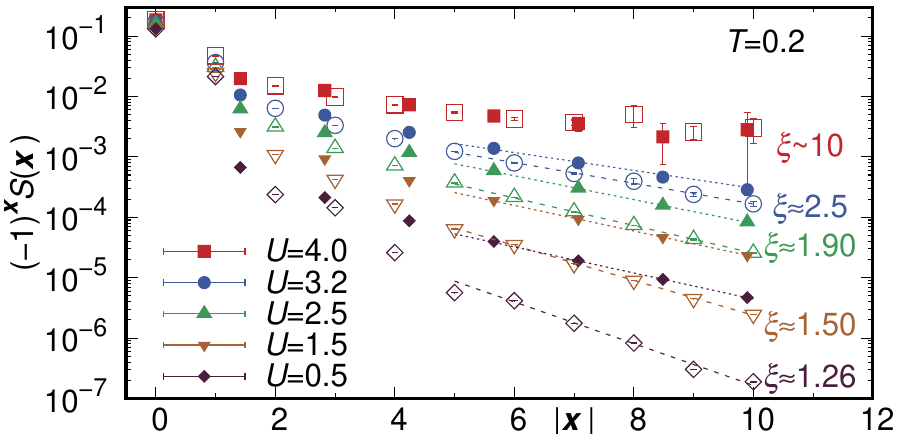}
	\caption{ Spin correlation function $S(\mathbf{x})$ for various $U$ at $T=0.2$. Solid (open) symbols correspond the diagonal (axial) direction. The linear fits define the correlation length $\xi$. }
	\label{fig:Sx_log}
\end{figure}

Magnetic signatures of the crossover are captured by the equal-time spin correlation function 
\begin{equation}
	S(\mathbf{x}-\mathbf{y}) = \langle s_z(\mathbf{x})s_z(\mathbf{y})\rangle,
	\label{}
\end{equation}
where $s_z(\mathbf{x})=(n_{\uparrow}(\mathbf{x}) - n_{\downarrow}(\mathbf{x}))/2$.
Its long-distance asymptotics define the AFM correlation length $\xi$ via $(-1)^\mathbf{x}S(\mathbf{x}) \propto \exp(-|\mathbf{x}|/\xi)$ in the $|\mathbf{x}| \to \infty$ limit. Figure~\ref{fig:Sx_log}, where it is plotted at $T=0.2$ and several $U$, shows that $\xi$ increases monotonically with $U$. In the metallic regime $\xi$ is of order of one lattice spacing. At $U \approx 2.5$ corresponding to $T^*_\mathrm{ch}(U)=0.2$, $\xi$ is only $\sim 2$. It should be noted, however, that the prefactor of $\exp(-|\mathbf{x}|/\xi)$ changes by orders of magnitude between $\xi \sim 1$ and $\xi \sim 2$, which is the leading effect when the correlations are so short ranged. The relatively short-range nature of spin correlations is typical for a strongly correlated nFL~\cite{wu2018, Walsh:2019uk, Arzhang:2019}. At lower $T$ and $U$ the crossover happens at increasingly larger values of $\xi$, which makes it increasingly mean-field-like, resulting in shrinking of the nFL region and vanishing differentiation between the nodal and antinodal values of the self-energy in Ref.~\cite{Fedor:2018vo}. As we enter the insulating regime, $\xi$ grows rapidly, becoming of order $10$ at $U=4$, $T=0.2$. 
At small $U$, the correlation length is anisotropic: $\xi$ and the values of $|S(\mathbf{x})|$ obtained from the asymptotics along the diagonal direction $\mathbf{x}=(x,x)$ are notably larger than those along the axis $\mathbf{x}=(0,x)$. The anisotropy is characteristic of the noninteracting limit and becomes negligible in the AFM regime at larger $U$ when $\xi\sim 10$. 

\begin{figure}[]
	\centering
	\includegraphics[width=0.45\textwidth]{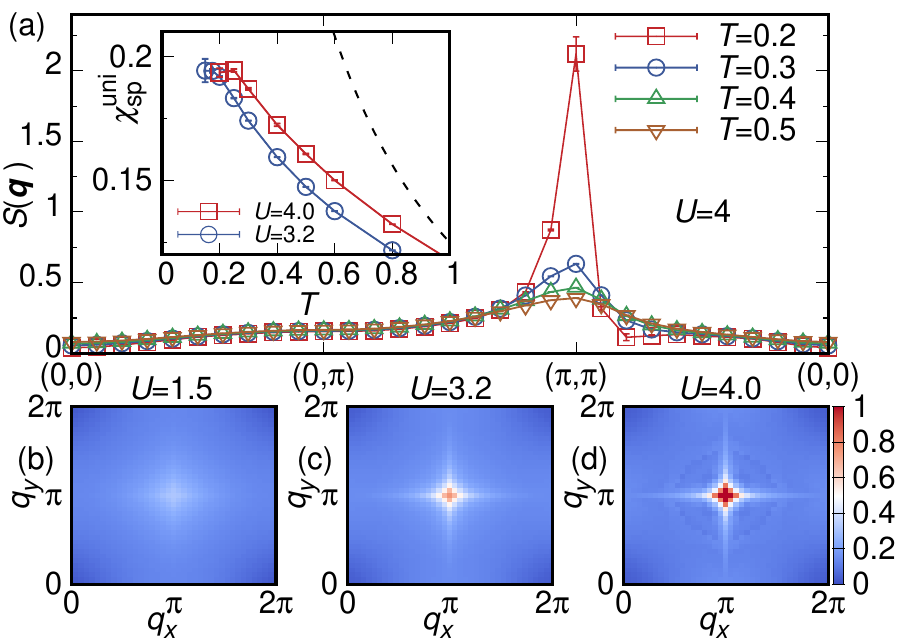}
	\caption{
		(a) Spin structure factor $S(\mathbf{q})$ along the high-symmetry momentum line.  
		Inset: the uniform static spin susceptibility as a function of temperature; the dashed line is Curie's law. 
		(b)-(d) $S(\mathbf{q})$ in the BZ for different $U$ at $T=0.2$.
	}
	\label{fig:Sq_BZ_Tdep}
\end{figure}

Development of the quasi-AFM state is seen in the magnetic structure factor $S(\mathbf{q}) = \sum_{\mathbf{x}} e^{-i\mathbf{q} \mathbf{x}} S(\mathbf{x})$. Upon increasing $U$ [Figs.~\ref{fig:Sq_BZ_Tdep}(b)-\ref{fig:Sq_BZ_Tdep}(d)] or lowering $T$ [Fig.~\ref{fig:Sq_BZ_Tdep}(a)], a sharp peak in $S(\mathbf{q})$ develops at $\mathbf{q}=\mathbf{Q}$, while the uniform [$\mathbf{q}=(0,0)$] structure factor is suppressed. At strong correlations, $S(\mathbf{q})$ exhibits intriguing anisotropy: Fig.~\ref{fig:Sq_BZ_Tdep}(a) shows suppression of $S(\mathbf{q})$ near the peak along the $(\pi,\pi)-(0,0)$ line with cooling, while the shape of $S(\mathbf{q})$ along $(\pi,0)-(\pi,\pi)$ is robust, forming a shoulder near the peak. Figure~\ref{fig:Sq_BZ_Tdep}(d), where $S(\mathbf{q})$ is plotted as a color map in the Brillouin zone (BZ), shows that the shoulder becomes pronounced at large $U$ and is restricted to a narrow line along $(\pi,0)-(\pi,\pi)$, resulting in the vertical cross shape surrounded by a near-circular halo. The isotropy of $\xi$ is due to dominance of the isotropic peak.

Being a direct signature of AFM correlations, the relative magnitude of the peak can be used to define the crossover to a quasi-AFM state~\cite{simkovic2017magnetic}. The $T$ dependence of the uniform static spin susceptibility $\chi_{\mathrm{sp}}^{\mathrm{uni}} = \chi_{\mathrm{sp}}(\mathbf{q}=\mathbf{0},i\omega_n=0) = \beta S(\mathbf{q}=0)$ offers a more visually compelling definition~\cite{Paiva:2010}. At high temperatures $\chi_{\mathrm{sp}}^{\mathrm{uni}}$ follows Curie's $1/T$ law. The renormalized classical regime of long-range AFM fluctuations, realized at low temperatures~\cite{Borejsza:2003gu,Borejsza:2004hh}, features a $\chi_{\mathrm{sp}}^{\mathrm{uni}}$ that increases with $T$ ~\cite{Hasenfratz:1993gt,Kim:1998xt}. Therefore, $\chi_{\mathrm{sp}}^{\mathrm{uni}}(T)$ must exhibit a maximum, seen in the inset of Fig.~\ref{fig:Sq_BZ_Tdep}, the location of which defines $T^*_{\mathrm{sp}}$~\cite{Paiva:2010}, \footnote{The recent results of the ladder dual fermion approximation~\cite{vanLoon:2018hm} are in quantitative agreement.}.

At $T^*_{\mathrm{sp}}$, $\xi$ is of order $10$, in consistency with its meaning as the crossover temperature to a quasi-AFM state. For $T \lesssim 0.25$, where the crossover is meaningful, $T^*_{\mathrm{sp}}$ coincides with $T^*_{\mathrm{n}}$. Thus, below $T^*_{\mathrm{sp}}\approx T^*_{\mathrm{n}}$ the whole FS is already gapped and the system is an insulator. The insulator boundary is well described by the N\'{e}el temperature of the dynamical mean-field theory (DMFT) $T^N_\mathrm{DMFT}$ \cite{Kunes:2011is} 
, although the qualitative behavior of $\chi_{\mathrm{sp}}^{\mathrm{uni}}(T)$ in this regime, which is controlled by extended spatial correlations, is not captured by DMFT~\cite{vanLoon:2018hm}. 

\begin{figure}[]
	\centering
	\includegraphics[width=0.45\textwidth]{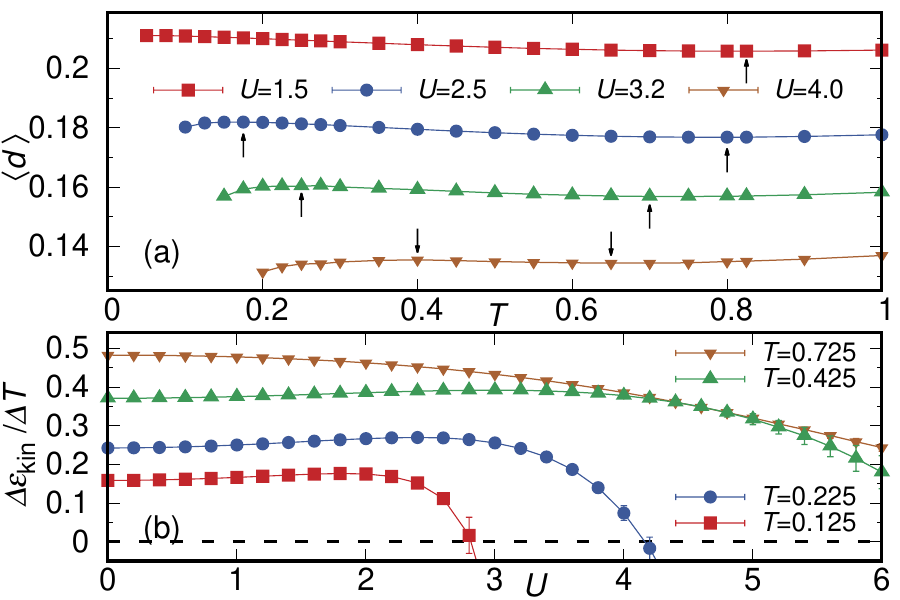}
	\caption{
		(a) Double occupancy as a function of $T$ for several $U$.
		(b) Numerical derivative of the kinetic energy with respect to $T$ as a function of $U$.
	}
	\label{fig:docc_dEpotdT}
\end{figure}

The nFL nature of the regime between $T^*_{\mathrm{ch}}$ and $T^*_{\mathrm{sp}}$ is further illustrated by energetics. Fig.~\ref{fig:docc_dEpotdT}(a) exemplifies $\langle d \rangle(T)$: for $U \lesssim 4.5$, it features a maximum at $T_{\mathrm{pot}}^{\mathrm{max}}(U)$ and a minimum at $T_{\mathrm{pot}}^{\mathrm{min}}(U)$ (marked by the arrows). $\langle d \rangle$ drops upon cooling ($\partial \langle d \rangle/\partial T>0$) from its high-$T$ asymptote $1/4$, reflecting formation of the local magnetic moment $\langle s_z^2 \rangle=(1-2\langle d \rangle)/4$. However, in the region $T_{\mathrm{pot}}^{\mathrm{max}}<T<T_{\mathrm{pot}}^{\mathrm{min}}$ [gray shading in Fig.~\ref{fig:phasediagram}(a)], it grows ($\partial \langle d \rangle/\partial T<0$), as expected for a half-filled metal~\cite{Georges:1993gz, Werner:2005hq, Kozik:2013ji,Fratino:2017jw}. Here, an adiabatic increase of $U$ leads to cooling~\cite{Werner:2005hq}---in analogy with the Pomeranchuk effect~\cite{Pomeranchuk:1950}---via the Maxwell relation $\partial \langle d \rangle/ \partial T = - \partial s/\partial U$, where $s$ is the entropy density. Thus, for $T_{\mathrm{pot}}^{\mathrm{max}}<T<T^*_{\mathrm{ch}}$, $\partial \kappa/\partial T>0$ as in an insulator but $\partial \langle d \rangle/\partial T<0$ as in a metal. The $T_{\mathrm{pot}}^{\mathrm{max}}$ line is notably above $T_{\mathrm{sp}}^*$: between these lines the local moment develops to support the extending AFM correlations, as above the N\'{e}el transition in three dimensions~\cite{Kozik:2013ji}. It is easily seen that without nonlocal density fluctuations (i.e. if $C_{\mathrm{nl}}=0$), the relation $T_{\mathrm{pot}}^{\mathrm{max}}<T^*_{\mathrm{ch}}$ would be reversed, which implies that they are important up until $T \sim 0.5$. 

The suppression of double occupancy by cooling and the corresponding reduction of the potential energy near $T^*_{\mathrm{sp}}$  is characteristic of a Slater AFM insulator~\cite{Rev:1951ib}. In this picture, it happens at the expense of a kinetic energy rise~\cite{Gull:2008fw,Rohringer:2016jt,Fratino:2017jw,vanLoon:2018hm}. Figure~\ref{fig:docc_dEpotdT}(b) shows the numerical derivative of $\varepsilon_\mathrm{kin}$ with respect to $T$ versus $U$. At low $T$, increasing $U$ leads to its sign change, $\Delta \varepsilon_\mathrm{kin}/\Delta T=0$ defining $T_{\mathrm{kin}}^*(U)$. Below $T_{\mathrm{kin}}^*$, $\Delta \varepsilon_\mathrm{kin}/\Delta T<0$, as it should in a Slater insulator. It is instructive that $T_{\mathrm{kin}}^* \approx T_{\mathrm{sp}}^*$, consistently with $T_{\mathrm{sp}}^*$ being the insulator boundary. For $T_{\mathrm{sp}}^*< T< T_{\mathrm{pot}}^{\mathrm{max}}$ the system is neither a metal nor insulator: both $\varepsilon_\mathrm{kin}$ and $\varepsilon_\mathrm{pot}$ are reduced upon cooling. The temperature $T_{\mathrm{kin}}^{\mathrm{max}}$ at which $\frac{\Delta \varepsilon_\mathrm{kin}}{\Delta T}(U)$ is maximal before dropping to change the sign marks the crossover between metallic and nFL behavior: it coincides with $T_{\mathrm{ch}}^*$ up to $T\sim 0.35$, whereas at higher $T$ the maximum is observed at decreasing $U$ and eventually disappears [Fig.~\ref{fig:phasediagram}(a)].

In summary, the transitional nFL behavior is a manifestation of the generic separation between the energy scales for fluctuations in different channels, which vanishes in the weak-coupling mean-field regime. It can be revealed in measurements of spin and charge correlations as well as energetics in the experimentally accessible range of parameters $T \lesssim 0.25$ and $ 2 \lesssim U \lesssim 4$. Since in this regime the extended nonlocal fluctuations in both channels play a crucial role, approaches limited to a finite system, either theoretical or experimental, require careful control of finite-size errors \footnote{We have checked that, e.g., $S(\mathbf{q}=[\pi, \pi])$ at $T=0.2$, $U=4$ computed on a $24 \times 24$ lattice with periodic boundary conditions underestimates the TDL value by $\sim 20\%$. In this regime, the crossover lines in Fig.~\ref{fig:phasediagram} typically shift beyond error bars for system sizes of order $10 \times 10$.
}. At $U \gtrsim 4$ cooling below $T^*_{\mathrm{ch}}$ brings the system from a thermal gas directly into the insulating regime, where the physics is mostly local, and eventually to the quasi-AFM state with $\xi \gtrsim 10$ at the much lower $T^*_{\mathrm{sp}}$. It is expected~\cite{Paiva:2001_specific_heat, Borejsza:2003gu} that at a large $U$, beyond the scope of Fig.~\ref{fig:phasediagram}, the nature of the insulating state will change from Slater to Mott-Heisenberg.

\section{acknowledgement}
The authors are grateful to the Precision Many-Body Group at UMass Amherst, where a part of this work was carried out, for hospitality. This work was supported by EPSRC through grant EP/P003052/1 and partially supported by the Simons Foundation as a part of the Simons Collaboration on the Many-Electron Problem. 
We are grateful to the UK Materials and Molecular Modelling Hub for computational resources, which is partially funded by EPSRC (EP/P020194/1).

\bibliography{ref.bib}

\end{document}